\documentclass[pra,showpacs]{revtex4}

\usepackage{amsfonts}
\usepackage{amssymb}
\usepackage{amsbsy}
\usepackage{bm}

\begin{document}


\title{A description of entanglement in
terms of quantum phase}

\author{Luis L. S\'anchez-Soto and Juli\'an Delgado}
\affiliation{Departamento de \'{O}ptica,
Facultad de Ciencias F\'{\i}sicas,
Universidad Complutense, 28040 Madrid, Spain}

\author{Andrei B. Klimov}
\affiliation{Departamento de F\'{\i}sica,
Universidad de Guadalajara, Revoluci\'on 1500,
44420 Guadalajara, Jalisco, M\'exico}

\author{Gunnar Bj\"ork}
\affiliation{Department of Microelectronics and
Information Technology, Royal Institute of Technology
(KTH), Electrum 229, SE-164 40 Kista, Sweden}

\date{\today}

\begin{abstract}
We explore the role played by the phase in an accurate
description of the entanglement of bipartite systems.
We first present an appropriate polar decomposition that
leads to a truly Hermitian operator for the phase of a single
qubit. We also examine the positive operator-valued measures
that can describe the qubit phase properties. When dealing
with two qubits, the relative phase seems to be a natural
variable to understand entanglement. In this spirit, we
propose a measure of entanglement based on this variable.
\end{abstract}


\maketitle

\bigskip

\section{Introduction}

The ability to create and manipulate quantum states
is a topic of increasing interest, with strong implications
in areas of futuristic technology such as quantum computing,
quantum cryptography, and quantum communications.

Entanglement, or nonlocal quantum correlations
between spacelike separated systems, constitutes the
essential resource for most, if not all, the applications
in quantum information~\cite{Nie01,Bou00}. In fact, the
idea of entanglement has proved to be one of the most fertile
and thought-generating properties of quantum mechanics
and much work has been devoted to understanding how it
can be quantified and manipulated~\cite{Ben96,Ved97a,Ved97,
Eis99,Vid99,Ben00,Vid00}

It is a general belief nowadays that bipartite pure-state
entanglement is almost completely understood~\cite{Tar01}
(although some questions are still open for the mixed state case).
This entanglement is usually introduced through quantum
states that violate the classical locality requirement
(i.e., violate the Clauser-Horne-Shimony-Holt
inequality). In fact, the mathematical tool underlying
this kind of treatments is the Schmidt bi-orthogonal
expansion~\cite{Per95,EkKn95}, since it allows us to write
any pure state shared by two parties in a canonical form,
where all the information about the nonlocal properties is
contained in the Schmidt coefficients. However, in physical
terms, entanglement is deeply linked to the superposition
principle, which also plays a key role in the puzzling
world of quantum interference. Thus, from this perspective,
it is hardly surprising that nonlocal correlations
can be interpreted in terms of well-defined
phase relations between parties~\cite{San00},
a point that goes almost unnoticed in the vast
literature on the subject, in spite of the fact that
phase is essential in understanding classical
correlation phenomena~\cite{Man95}.

Perhaps the most spectacular application of
entanglement is the quantum computer, which
would allow an exponential increase of computational
speed for certain problems, once it is realized. The
core of quantum computing are the logic gates~\cite{Ek95}
(it is known that any quantum computation can
be reduced to a sequence of universal two qubit logic
gates and one qubit local operation~\cite{Vin95}). The
realization of these logic gates in trapped ions~\cite{Mon95},
cavity QED~\cite{Tur95,Rau99} and NMR~\cite{Chu98,Jon98}
have already been possible, showing the practical realization
of two equivalent kinds of universal two qubit logic
gates: a quantum controlled not gate  and a
quantum phase gate  that differ from each
other only by local operations~\cite{Sol01}.
The main problem these systems face is decoherence:
it is essential to keep coherence of the qubits themselves
and among them.

The moral we wish to extract from the previous
discussion is that the concepts of phase and coherence
for qubits are ubiquitous in the modern parlance of
quantum information. However, the notions of phase
for a qubit and of relative phase between qubits is
loosely used in the literature, mainly because there is a
lack of a clear prescription for dealing with such variables
in the quantum world. Even worse, sometimes the concepts
employed in this field are misleading. Therefore,
a thorough explanation of the techniques needed to
characterize the phase of qubit states will be of
relevance to workers in the various diverse experimental
fields currently under consideration for quantum
computing technology. This is the main goal of this paper.

\section{Phase for a single qubit}

The system we wish to study lives in a two-dimensional
Hilbert space spanned by two states, we shall denote by $|0
\rangle $ and $| 1 \rangle $. In this Hilbert space, the operators
\begin{eqnarray}
& \hat{S}_+ = | 1 \rangle \langle 0 | ,
\qquad
\hat{S}_- = | 0 \rangle \langle 1 | , &
\nonumber \\
& \hat{S}_z = \frac{1}{2}
(| 1 \rangle \langle 1 | - | 0 \rangle \langle 0 |) ,
\end{eqnarray}
together with the identity $\hat{I}$, form a complete
set of linearly independent observables. They satisfy
the commutation relations
\begin{equation}
\label{ccr}
[ \hat{S}_z, \hat{S}_\pm ] = \pm \hat{S}_\pm ,
\qquad
[ \hat{S}_+, \hat{S}_- ] =  2 \hat{S}_z ,
\end{equation}
which are distinctive from the su(2) algebra that describes
angular momentum in quantum mechanics.

Loosely speaking, a two-state system is called a quantum
bit or qubit~\cite{Schu94}, in direct analogy with the
classical bit of information (which consists in two
distinguishable states of some system). Unlike the classical
bit, a qubit can be in a superposition of the form
\begin{equation}
\label{qubit}
| \Psi \rangle = \sin(\vartheta/2) | 0 \rangle +
e^{i \varphi} \ \cos(\vartheta/2) | 1 \rangle .
\end{equation}
This corresponds to a 1/2 angular-momentum system, and we
know~\cite{Cohen90} that for this particular case $\hat{\mathbf{S}}
= \hat{\bm{\sigma}}/2$, $\hat{\bm{\sigma}}$ being the Pauli
matrices. Then it is easy to work out that the mean values
$s_j = \langle \Psi | \hat{S}_j | \Psi \rangle$ are given by
\begin{eqnarray}
s_x & = & \sin \vartheta \ \cos \varphi , \nonumber \\
s_y & = & \sin \vartheta \ \sin \varphi , \nonumber \\
s_z & = & \cos \vartheta ,
\end{eqnarray}
where $\hat{S}_\pm = (\hat{S}_x \pm i \hat{S}_y)$.
This would support the naive belief that, when viewed in
the Bloch sphere $s_x^2 + s_y^2 + s_z^2 =1$, the
parameter $\varphi$ appears as the phase angle associated
with the qubit and that it is canonically conjugate to
$s_z$~\cite{Luis00}. Therefore, there is a widespread
usage of dealing with this qubit phase $\varphi$ as
a state parameter instead of a quantum variable,
as one could expect from the very basic principles
of quantum mechanics~\cite{Per95}.

To gain further insight in this point, let us note that
\begin{equation}
\label{s-}
s_- =  \langle \Psi | \hat{S}_- | \Psi \rangle =
\sin \vartheta \ e^{i \varphi} ,
\end{equation}
so it is clear that the parameter $\varphi$ can
be obtained through the decomposition of $s_-$ in
terms of modulus and phase. Obviously, it is tempting
to pursue this simple picture by taking
into account that, at the operator level, the equivalent
to the decomposition in terms of modulus and phase
is a polar decomposition. Thus, it seems appropriate
to define the operator counterpart of Eq.~(\ref{s-})
by
\begin{equation}
\label{despolar}
\hat{S}_- =  \sqrt{\hat{S}_- \hat{S}_+} \ \hat{E} .
\end{equation}
Here the unitary operator $\hat{E}$ represents the complex
exponential of the qubit phase; i.e., $\hat{E}=
e^{i \hat{\Phi}}$, where $\hat{\Phi}$ is the Hermitian
operator for the phase. The use of the polar decomposition
of su(2) was pioneered by L\'{e}vy-Leblond~\cite{Lev73},
and worked out from the mathematical viewpoint in
Refs.~\cite{Vou90} and \cite{Ell90}. Its application
to the proper description of the atomic-dipole phase
in quantum optics was fully developed in Ref.~\cite{phasJCM}.

It is not difficult to work out that the unitary solution of
Eq.~(\ref{despolar}) is
\begin{equation}
\hat{E} = | 0 \rangle \langle 1 | + e^{i \varphi_0} \
 | 1 \rangle \langle 0 | ,
\end{equation}
where $\varphi_0$ is some arbitrary phase that
corresponds to a matrix element undefined by
Eq.~(\ref{despolar}) and that appears due to the
unitarity requirement. The main features of this
operator are largely independent of $\varphi_0$,
however for the sake of concreteness we can
make a definite choice. For instance, according to
Eq.~(\ref{s-}), the complex conjugation of the qubit
wave function $| \Psi \rangle $ should reverse the
sign of  $\hat{\Phi}$~\cite{OC97}. This leads to
the condition $e^{i \varphi_0} = -1$, and therefore
\begin{equation}
\hat{E} = | 0 \rangle \langle 1 | - | 1 \rangle \langle 0 | ,
\end{equation}
whose eigenvectors are
\begin{equation}
| \varphi_\pm \rangle = \frac{1}{\sqrt{2}}
( | 0 \rangle \pm i  | 1 \rangle ) .
\end{equation}
It is worth mentioning that these states have
been exploited in recent proposals of covariant
cloning~\cite{Bru00,Ari01,Cer02}.

To any smooth function $F(\varphi)$ we can
associate the operator
\begin{equation}
F(\hat{\Phi}) = \sum_\pm | \varphi_\pm \rangle \
F (\varphi_\pm) \ \langle \varphi_\pm | ,
\end{equation}
where the sum runs over the two possible eigenvalues.
The mean value of this operator function can be computed as
\begin{equation}
\label{Fmean}
\langle F(\hat{\Phi}) \rangle =
\sum_\pm F (\varphi_\pm) P(\varphi_\pm ) ,
\end{equation}
where $P(\varphi_\pm)$ is the probability distribution
\begin{equation}
\label{Pphi}
P(\varphi_\pm) = \mathrm{Tr} \left [ \hat{\varrho} \
|\varphi_\pm \rangle \langle \varphi_\pm | \right ] ,
\end{equation}
for any state described by the density matrix $\hat{\varrho}$. Two
important remarks seem pertinent. First, note that $\hat{E}$ is
proportional to $\hat{S}_y$. Also, because $\hat{E}^\dagger
= - \hat{E}$, we have $\cos \hat{\Phi} = 0$, which is,
certainly, a rather pathological behavior. This is caused
by the small dimension of the system space as such strong
relations no longer hold for dimensions greater than two.
In other words, the two-dimensional Hilbert space where the
qubit lives is not large enough to distinctly accommodate all
different variables.

Second, and perhaps more striking, is that the qubit phase
can take only two values, $\pm \pi/2$, due also to the
dimension of the Hilbert space. While this statement
seems rather reasonable when dealing with spin
systems, it is scarcely recognized when dealing
with qubits. Note the clear distinction  between
the state parameter $\varphi$ in Eq.~(\ref{qubit}) and
the outcome of a  measurement of the qubit phase:
the former is continuous, while the later is discrete
and binary. We shall elaborate on this point in
next Section.

\section{Positive-operator valued measures
for qubit phase}

Given the singular behavior exhibited by
the description of qubit phase in terms of
a Hermitian operator, one may think preferable
to use a positive-operator valued measure
(POVM) taking continuous values in a
$2 \pi$ interval. Additionally, this formalism
can include also fuzzy generalizations of
the ideal phase description provided by $ |\varphi_\pm \rangle
\langle \varphi_\pm |$, as in Eq.~(\ref{Pphi}).
To keep the discussion as self-contained as possible,
we briefly recall that a POVM~\cite{Hel76,Sha91} is
a set of linear operators $\hat{\Delta} (\varphi)$ furnishing
the correct probabilities in any measurement
process through the fundamental postulate that
\begin{equation}
P(\varphi ) = \mathrm{Tr} [
\hat{\varrho} \ \hat{\Delta} (\varphi) ].
\end{equation}
The real valuedness, positiveness, and normalization of
$P(\varphi)$ impose
\begin{equation}
\label{condpom}
\hat{\Delta}^\dagger (\varphi) = \hat{\Delta} (\varphi) ,
\quad
\hat{\Delta} (\varphi) \ge 0 ,
\quad
\int_{2 \pi} d\varphi \ \hat{\Delta} (\varphi) = \hat{I},
\end{equation}
where the integral extends over any $2 \pi$ interval
of the form $(\varphi_0, \varphi_0 + 2 \pi)$, $\varphi_0$
being a fiducial or reference phase. Note that,
in general, $\hat{\Delta} (\varphi)$ are not
orthogonal projectors like in the standard von Neumann
measurements described by selfadjoint operators.

In addition to these basic statistical conditions,
some other requirements must be imposed
to ensure that $\hat{\Delta} (\varphi)$ provides
a meaningful description of the phase as a canonically
conjugate variable with respect $\hat{S}_z$ (even
in the sense of a weak Weyl relation~\cite{Gra95}).
Then, we require~\cite{Leo95}
\begin{equation}
\label{req1}
e^{i \varphi^\prime \hat{S}_z} \
\hat{\Delta} (\varphi) \ e^{- i \varphi^\prime \hat{S}_z}
= \hat{\Delta} (\varphi + \varphi^\prime) ,
\end{equation}
which reflects nothing but the basic feature
that a phase shifter is a phase-distribution
shifter.

We must also take into account that a shift in
$\hat{S}_z$ should not change the phase distribution.
Therefore, we require as well
\begin{equation}
\label{req2}
\hat{E}  \ \hat{\Delta} (\varphi) \ \hat{E}^\dagger
= \hat{\Delta} (\varphi ) ,
\end{equation}
which, loosely speaking, is the physical translation
of the fact that phase should be complementary to
the action variable $\hat{S}_z$.

In Ref.~\cite{phasJCM} it has been shown that for the
case of an angle variable the most general POVM
fulfilling these two properties is of the form
\begin{equation}
\label{Dgen}
\hat{\Delta}_\gamma (\varphi) = \frac{1}{2 \pi}
( \hat{I} + \gamma e^{i \varphi} \hat{S}_+ +
\gamma e^{- i \varphi} \hat{S}_- ) ,
\end{equation}
where $\gamma \le 1$ is a real number (otherwise
its argument can be included in the definition
of $\varphi$) whose physical meaning will be
elucidated soon. The phase distribution $P(\varphi)$
induced by this POVM is of the form
\begin{equation}
\label{Pcgen}
P (\varphi) = \frac{1}{2 \pi}
( 1 + c e^{i \varphi} + c^\ast e^{- i \varphi} ) ,
\end{equation}
with $c= \langle 0 |\hat{\varrho} | 1 \rangle \
\gamma $. This is a remarkably simple result
that holds for any mixed state of the system.

This form allows us to examine some very
general properties of the qubit phase. First,
note that the information $P(\varphi)$ conveys
goes beyond what would strictly be the phase.
Indeed, since
\begin{eqnarray}
\langle \hat{S}_x \rangle & = &
\frac{1}{\gamma} \int_{2\pi} d\varphi \
\cos \varphi \ P(\varphi), \nonumber \\
\langle \hat{S}_y \rangle & = &
\frac{1}{\gamma} \int_{2\pi} d\varphi \
\sin \varphi \ P(\varphi) ,
\end{eqnarray}
and $\hat{S}_x^2 = \hat{S}_y^2 = \hat{I}/4$,
$P(\varphi)$ contains the complete statistics of
$\hat{S}_x$ and $\hat{S}_y$. In fact, it
contains the whole statistics of the qubit and
not only of its phase.

Suppose now that we have the qubit described in
terms of two different POVMs with $\gamma_1$ and
$\gamma_2$, such that $ \gamma_1 < \gamma_2$. If
one takes the dispersion $D$ as a measure
of the phase uncertainty of the state, we have
\begin{equation}
D_j^2 = 1 - \left |
\int_{2\pi} d\varphi \ e^{ i \varphi}
P_j (\varphi) \right |^2 = 1 - | c_j|^2 ,
\end{equation}
and then, $D_1 \ge D_2$. This shows that $P_1 (\varphi)$
is always broader than $P_2 (\varphi)$ when
$ \gamma_1 < \gamma_2$. Moreover, one can
also check that
\begin{equation}
\hat{\Delta}_{\gamma_1} (\varphi) =
\frac{1}{2 \pi} \int_{2 \pi}
d\varphi^\prime \ \left [
1 + \frac{\gamma_1}{\gamma_2}
e^{i (\varphi - \varphi^\prime)}
+ \frac{\gamma_1^\ast}{\gamma_2^\ast}
e^{- i (\varphi - \varphi^\prime)} \right ] \
\hat{\Delta}_{\gamma_2}(\varphi^\prime) .
\end{equation}
Therefore, both POVMs contain the same information about
the qubit: if one of them is known, the other one can be
directly obtained.

A relevant feature of this approach is that
it provides a qubit phase where any value of
$\varphi$ is allowed. However, we wish to emphasize
that this continuous range of variation is not effective
in the sense that the values of $P(\varphi)$ at
every point $\varphi$ cannot be independent, and we
can find relations between them irrespective of the
qubit state. In other words, $\hat{\Delta} (\varphi)$
cannot be linearly independent because the qubit
Hilbert space is two dimensional and the algebra of
operators acting on that Hilbert space is four dimensional.

This can be stated in a slightly different way: given
the general form of $P(\varphi)$ in Eq.~(\ref{Pcgen}),
the complex parameter $c$ can be determined by the
value of $P(\varphi)$ at two $\varphi $ points.
Nevertheless, more manageable expressions appear
if we use three points instead of
two, such as $\varphi_r = 2 \pi r/3$ ($r = -1, 0, 1$).
After some calculations we get
\begin{equation}
c = \frac{2 \pi}{3} \sum_{r= 0, \pm 1}
P(\varphi_r) e^{- i \varphi_r} ,
\end{equation}
which allows us to express $P(\varphi)$ as
\begin{equation}
P(\varphi) = \frac{1}{3} \sum_{r,s = 0, \pm 1}
P(\varphi_r) e^{i s (\varphi - \varphi_r)} ,
\end{equation}
and so the knowledge of the three values $P(\varphi_r)$
gives $P(\varphi)$ at any other point $\varphi$.

This effective discreteness allows us to compute
the mean values of any function $F(\varphi)$ in a
way very similar to Eq.~(\ref{Fmean}). Indeed,
one has
\begin{equation}
\langle F(\varphi) \rangle = \frac{2 \pi}{3}
\sum_{r = 0, \pm 1} \tilde{F} (\varphi_r) P(\varphi_r) ,
\end{equation}
where $\tilde{F}$ is related to $F$ by
\begin{eqnarray}
\label{Fprop}
& \displaystyle
\int_{2 \pi} d\varphi \ e^{i k \varphi} \tilde{F}(\varphi) =
\int_{2 \pi} d\varphi \ e^{i k \varphi} F(\varphi) ,
\qquad
k = 0, \pm 1, &
\nonumber \\
& & \nonumber \\
& \displaystyle
\int_{2 \pi} d\varphi \ e^{i k \varphi} \tilde{F}(\varphi) = 0 ,
\qquad
|k| = \pm 2, \pm 3, \ldots . &
\end{eqnarray}
This last equation guarantees that for the function
$e^{i k \varphi}$ we have $\langle e^{i k \varphi}
\rangle = 0$, as it should be.
Moreover, Eqs.~(\ref{Fprop}) imply that
\begin{equation}
\int_{2 \pi} d\varphi \ P(\varphi) \tilde{F}(\varphi) =
\int_{2 \pi} d\varphi \ P(\varphi) F(\varphi)  ,
\end{equation}
for any $P(\varphi)$. We conclude then that
discreteness is inevitably at the heart of the qubit
phase.

Finally, we shall consider two particular examples
of POVMs that are admissible to describe the qubit
phase. The first one consists in
\begin{equation}
\hat{\Delta}_{\mathrm{SG}} (\varphi) =
| \varphi \rangle \langle \varphi | ,
\end{equation}
where
\begin{equation}
| \varphi \rangle= \frac{1}{\sqrt{2 \pi}}
(| 0 \rangle +e^{i \varphi} | 1 \rangle ) .
\end{equation}
This is a finite-dimensional translation of the
POVM generated by the Susskind-Glogower phase
states in the Fock space of a harmonic
oscillator~\cite{SG64,revphas}. One can
check that it is of the general form (\ref{Dgen}) with
$\gamma = 1$. We think that the definition of this POVM is
rather reasonable: while the operator $\hat{E}$ selects an
orthogonal basis from the set $|\varphi \rangle$, this POVM,
on the contrary, does not privilege any $|\varphi \rangle$
and all of them play the same role.

In the case of a single-mode quantum field, a widely used
conception of phase is based on examining quasiprobability
distributions in phase space~\cite{Per91,Sch01}. Among them,
one of the most interesting and studied comes from the $Q$ function.
A POVM for the field phase can be defined then in terms
of radial integration, much in the spirit of the classical
conception. The natural translation of this procedure to
our qubit problem involves the use of SU(2) coherent
states for a 1/2 angular momentum~\cite{Per86}
\begin{equation}
| \vartheta, \varphi \rangle =
\sin(\vartheta/2) | 0 \rangle +
 e^{i \varphi} \ \cos(\vartheta/2) | 1 \rangle .
\end{equation}
and the $Q$-function they define as
\begin{equation}
Q ( \vartheta, \varphi ) = \frac{1}{2 \pi}
\mathrm{Tr} [ \hat{\varrho} \
| \vartheta, \varphi \rangle \langle | \vartheta, \varphi | ] .
\end{equation}
This $Q$ function can be regarded as a natural
probability distribution in the associated qubit phase space,
which is the manifold of SU(2). Now, we can define a POVM for
$\varphi$ by a marginal integration over $\vartheta$; that is,
\begin{equation}
\hat{\Delta}_Q (\varphi ) = \frac{1}{2 \pi} \int_0^\pi
d \vartheta \ \sin \vartheta \
| \vartheta, \varphi \rangle \langle \vartheta, \varphi | ,
\end{equation}
which corresponds to Eq.~(\ref{Dgen}) with $\gamma =
\pi/4.$

Note that we have focused on the case of a single
qubit. The generalization of this formalism to a
collection of $N$ identical qubits is straightforward:
the operators $\hat{S}_+$ and $\hat{S}_+$ constitute
then a $(N+1)$-dimensional representation of su(2)~\cite{Per95}
and, in consequence, the qubit phase takes $N+1$
distinct values. In terms of POVMs, the discreteness
is also present but now we need to know the value of
$P (\varphi)$ at  $N+1$ independent points. In other
words, the general form of $P (\varphi)$ involves only
$N+1$ different frequencies, and the Fourier series
can be inverted from the knowledge of $P (\varphi)$
in $N+1$ points.

\section{Relative phase for two qubits}

Let us assume now that we have two such qubits,
which we shall label by the subscripts A and B (in the
language of quantum information, the first qubit
belongs to Alice and the second one belongs to Bob).
It seems natural to introduce the exponentials of the
phase sum $\hat{E}_{(+)}$ and phase difference
$\hat{E}_{(-)}$ by the unitary operators
\begin{equation}
\hat{E}_{(+)} = \hat{E}_{\mathrm{A}}
\hat{E}_{\mathrm{B}} ,
\qquad
\hat{E}_{(-)} = \hat{E}_{\mathrm{A}}
\hat{E}_{\mathrm{B}}^\dagger .
\end{equation}
We have used the subscripts $(+)$ and $(-)$ for sum
and difference to distinguish clearly from the symbols of
ladder operators. We also introduce the operators
(for simplicity we drop henceforth the subscript $z$ from
$\hat{S}_{(+)}$ and $\hat{S}_{(-)}$, since there is no
risk of confusion)
\begin{equation}
\hat{S}_{(+)} =
\hat{S}_{z, \mathrm{A}} + \hat{S}_{z, \mathrm{B}} ,
\qquad
\hat{S}_{(-)} =
\hat{S}_{z, \mathrm{A}} - \hat{S}_{z, \mathrm{B}} ,
\end{equation}
that satisfy the commutation relations
\begin{eqnarray}
& [\hat{E}_{(+)}, \hat{S}_{(+)} ]  = \hat{E}_{(+)} ,
\qquad
[\hat{E}_{(-)}, \hat{S}_{(-)} ]  = \hat{E}_{(-)} ,
\nonumber \\
& & \nonumber \\
& [\hat{E}_{(+)}, \hat{S}_{(-)} ]  = 0 ,
\qquad
[\hat{E}_{(-)}, \hat{S}_{(+)} ] = 0 . &
\end{eqnarray}
so $\hat{E}_{(+)}$ ($\hat{E}_{(-)}$) is canonically
conjugate to $\hat{S}_{(+)}$ ($\hat{S}_{(-)}$).
Note also that the vectors $| \varphi_{\mathrm{A}},
\varphi_{\mathrm{B}} \rangle = | \varphi_{\mathrm{A}}
\rangle \otimes | \varphi_{\mathrm{B}} \rangle $ are
simultaneous eigenvectors of $\hat{E}_{(+)}$ and
$\hat{E}_{(-)}$, with eigenvalues $e^{i \varphi_+} =
e^{i (\varphi_{\mathrm{A}} + \varphi_{\mathrm{B}})}$
and $e^{i \varphi_-} = e^{i (\varphi_{\mathrm{A}} -
\varphi_{\mathrm{B}})}$, respectively.

The previous definition of phase sum and difference
seems appropriate because it is in accordance with the
algebra of complex numbers. However, due to its periodic
character, adding and subtracting phases must be
done with some care~\cite{Bar90,PDPD}. Since each
individual phase is expressed in a $2\pi$ range, the
eigenvalue spectra of the sum and difference operators
have widths of $4 \pi$, and this is not compatible with
the idea that a phase variable (even if it is a
phase sum or difference) must be $2 \pi $ periodic.
Thus, we must devise a way to cast the phase sum and
difference into the $2 \pi $ range.

This problem can be traced back to the fact that
while $(\hat{E}_{\mathrm{A}}, \hat{E}_{\mathrm{B}})$,
$(\hat{S}_{z, \mathrm{A}}, \hat{S}_{z, \mathrm{B}})$
or $(\hat{S}_{(+)}, \hat{S}_{(-)}) $, are complete
sets of commuting operators, this is not true for
$(\hat{E}_{(+)}, \hat{E}_{(-)})$, since the vectors
$| \varphi_{\mathrm{A}}, \varphi_{\mathrm{B}} \rangle$
and $| \varphi_{\mathrm{A}} + \pi ,
\varphi_{\mathrm{B}} + \pi \rangle$ have the same
phase sum and difference. In consequence, another
commuting operator must be considered to describe
the system. In Ref.~\cite{angle}, dealing with
the problem of angle sum and difference, it has
been proposed using the operator
\begin{equation}
\hat{V} = e^{2 i \pi \hat{S}_{(+)}} ,
\end{equation}
which commutes with $\hat{E}_{(+)}$ and
$\hat{E}_{(-)}$:
\begin{equation}
[\hat{E}_{(+)}, \hat{V} ] = [\hat{E}_{(-)}, \hat{V} ]  = 0.
\end{equation}
Therefore, $(\hat{E}_{(+)}, \hat{E}_{(-)}, \hat{V})$, is
a complete set of commuting operators, whose associated
basis is
\begin{equation}
\label{def}
| \varphi_+, \varphi_-, v \rangle =
\frac{e^{i v \varphi_{\mathrm{A}}}}{2}
\left [| \varphi_{\mathrm{A}},
\varphi_{\mathrm{B}} \rangle + (-1)^v
| \varphi_{\mathrm{A}} +
\pi, \varphi_{\mathrm{B}} + \pi \rangle
\right ] ,
\end{equation}
with $v = 0, 1$ and
\begin{equation}
\varphi_{\mathrm{A}} = \frac{1}{2}
(\varphi_+ + \varphi_-) ,
\qquad
\varphi_{\mathrm{B}} = \frac{1}{2}
(\varphi_+ - \varphi_-) .
\end{equation}
The complex exponential in the definition~(\ref{def})
has been introduced for convenience, in order to get
the same expression $| \varphi_+, \varphi_-, v \rangle $
when $ \varphi_{\mathrm{A}}$ and $\varphi_{\mathrm{B}}$
are replaced by $ \varphi_{\mathrm{A}} + \pi$ and
$ \varphi_{\mathrm{B}} + \pi$.
Then, the action of $\hat{V}$ in this basis is
\begin{equation}
\hat{V} | \varphi_+, \varphi_-, v \rangle =
(-1)^v | \varphi_+, \varphi_-, v \rangle ,
\end{equation}
and we have the resolution of the identity
\begin{equation}
\hat{I} = \sum_{v = 0, 1} \int_{2 \pi} \int_{2 \pi} d \varphi_+ d \varphi_- \
| \varphi_+, \varphi_-, v \rangle
\langle \varphi_+, \varphi_-, v | .
\end{equation}
The proper joint probability distribution $\mathcal{P}$
(cast into a $2 \pi$ range) for the phase sum
and difference associated with a system state
$\hat{\varrho}$ is
\begin{equation}
\mathcal{P}(\varphi_+, \varphi_-) =
\sum_{v = 0, 1} \langle \varphi_+, \varphi_-, v
| \hat{\varrho} | \varphi_+, \varphi_-, v \rangle ,
\end{equation}
which is nothing but the sum of the contributions
from each value of $v$.

We can express $\mathcal{P}(\varphi_+, \varphi_-) $
in terms of the probability distribution for the individual
phases $P (\varphi_{\mathrm{A}},
\varphi_{\mathrm{B}}) = \langle \varphi_{\mathrm{A}},
\varphi_{\mathrm{B}} |\hat{\varrho} |
\varphi_{\mathrm{A}}, \varphi_{\mathrm{B}} \rangle$
in the form
\begin{eqnarray}
\label{law}
\mathcal{P}(\varphi_+, \varphi_-) & = &
\frac{1}{2} \{ P \left [(\varphi_+ + \varphi_-)/2,
(\varphi_+ - \varphi_-)/2 \right ] \nonumber \\
& + & P \left [(\varphi_+ + \varphi_-)/2 + \pi,
(\varphi_+ - \varphi_-)/2 + \pi \right ] \} . \nonumber \\
\end{eqnarray}
Another equivalent procedure is to note that we must
get the same mean values for any periodic function of
the phase sum and difference whether we use the
variables $(\varphi_+, \varphi_-) $ or
$(\varphi_{\mathrm{A}}, \varphi_{\mathrm{B}}) $,
which translates into
\begin{eqnarray}
& \displaystyle
\int_{2 \pi} \int_{2 \pi} d\varphi_+ d\varphi_- \
e^{i k \varphi_+} e^{i l \varphi_-} \
\mathcal{P}(\varphi_+, \varphi_-) &  \nonumber \\
& \displaystyle = \int_{2 \pi} \int_{2 \pi} d\varphi_{\mathrm{A}}
d\varphi_{\mathrm{B}} \
e^{i k (\varphi_{\mathrm{A}} + \varphi_{\mathrm{B}})}
e^{i l (\varphi_{\mathrm{A}} - \varphi_{\mathrm{B}})} \
P(\varphi_{\mathrm{A}}, \varphi_{\mathrm{B}}) . & \nonumber \\
\end{eqnarray}
Since $\mathcal{P}(\varphi_+, \varphi_-) $ and
$P(\varphi_{\mathrm{A}}, \varphi_{\mathrm{B}}) $ are
$2 \pi$-periodic functions, these equalities determine
$\mathcal{P}(\varphi_+, \varphi_-) $ completely,
as can be shown by using a simple Fourier analysis~\cite{angle}.

We can now generalize the transformation law
(\ref{law}) to any POVM. It is clear that the joint
probability distribution function
$P(\varphi_{\mathrm{A}}, \varphi_{\mathrm{B}})$ arises from
$\hat{\Delta}(\varphi_{\mathrm{A}}, \varphi_{\mathrm{B}})$
defined by
\begin{equation}
\hat{\Delta} (\varphi_{\mathrm{A}},
\varphi_{\mathrm{B}}) = \hat{\Delta}_{\gamma_\mathrm{A}}
(\varphi_{\mathrm{A}}) \otimes
\hat{\Delta}_{\gamma_\mathrm{B}} (\varphi_{\mathrm{B}}) ,
\end{equation}
and then the use of (\ref{law}) leads to the following POVM
for the phase sum and difference cast into a $2 \pi$ range
\begin{eqnarray}
\label{Laque}
\hat{\Lambda} (\varphi_+, \varphi_-) & = &
\frac{1}{2} \{ \hat{\Delta} \left [(\varphi_+ + \varphi_-)/2,
(\varphi_+ - \varphi_-)/2 \right ] \nonumber \\
& + & \hat{\Delta} \left [(\varphi_+ + \varphi_-)/2 + \pi,
(\varphi_+ - \varphi_-)/2 + \pi \right ] \} . \nonumber \\
\end{eqnarray}

When focusing on the phase difference, the associated
POVM is defined by
\begin{equation}
\hat{\Lambda}(\varphi_-) =\int_{2 \pi} d\varphi_+ \
\hat{\Lambda}(\varphi_+, \varphi_-) ,
\end{equation}
which is equivalent to
\begin{equation}
\label{Lambda}
\Lambda(\varphi_-) = \int_{2 \pi} d\varphi^\prime \
\hat{\Delta}_{\gamma_\mathrm{A}} (\varphi_- + \varphi^\prime )
\hat{\Delta}_{\gamma_\mathrm{B}} (\varphi^\prime ) .
\end{equation}
This equation allows us to provide an alternative
approach to the fuzzy description of phase~\cite{Hel74}.
If we consider that the density operator factorizes,
$\hat{\varrho}_{\mathrm{A}} \otimes \hat{\varrho}_{\mathrm{B}}$,
the phase difference can be regarded as a measure
of the phase $\varphi_{\mathrm{A}}$ relative to a given
reference state described by $\hat{\varrho}_{\mathrm{B}}$.

\section{Degree of entanglement for two qubits}

Various measures of entanglement have been
proposed~\cite{Jae93,Shi95,Woo97,Run01,Abo01,Man02}
so far, each one having its own merits and emphasizing
different aspects of the phenomenon. We stress, however,
that a closer examination reveals that much
of these seemingly unconnected notions are
actually identical~\cite{Abo01}.

Given the distinguished role assigned in this
paper to the relative phase, it seems almost
compulsory to discuss a possible measure of
entanglement within this framework. For
simplicity, we shall restrict our attention
to the case of two qubits.

Let us first recall some previous well-established
definitions. A bipartite state $| \Psi_{\mathrm{A}
\times \mathrm{B}} \rangle$ is said to be
factorizable if it can be factored into a product
$| \Psi_{\mathrm{A} \times \mathrm{B}} \rangle =
| \Psi_{\mathrm{A}} \rangle \otimes
| \Psi_{\mathrm{B}} \rangle,$ where
$| \Psi_{\mathrm{A}} \rangle \in
\mathcal{H}_{\mathrm{A}}$ and $| \Psi_{\mathrm{B}}
\rangle \in \mathcal{H}_{\mathrm{B}}$,
and $\mathcal{H}_{\mathrm{A}}$ and
$\mathcal{H}_{\mathrm{B}}$ are the Hilbert
spaces of the individual qubits. An
entangled state is one for which this is not
possible. A maximally entangled bipartite
state  $| \Psi_{\mathrm{max}} \rangle$ satisfies
the conditions
\begin{equation}
\label{max}
\mathrm{Tr}_{\mathrm{A}} (
| \Psi_{\mathrm{max}}  \rangle
\langle \Psi_{\mathrm{max}} | ) =
\frac{1}{2} \hat{I}_{\mathrm{B}} ,
\;
\mathrm{Tr}_{\mathrm{B}} (
| \Psi_{\mathrm{max}}  \rangle
\langle \Psi_{\mathrm{max}} | ) =
\frac{1}{2} \hat{I}_{\mathrm{A}} ,
\end{equation}
where $\mathrm{Tr}_{\mathrm{A}}$
and $\mathrm{Tr}_{\mathrm{B}}$
stand for tracing over the subspaces
$\mathcal{H}_{\mathrm{A}}$ and
$\mathcal{H}_{\mathrm{B}}$, respectively.
Equation~(\ref{max}) implies that
each subsystem, when considered alone,
is in a maximally mixed state, although
the state of the system as a whole is
pure.

The general bipartite state of two qubits
may be expanded in the $\{ |0 \rangle,
| 1 \rangle \}$ bases of $\mathcal{H}_{\mathrm{A}}$
and $\mathcal{H}_{\mathrm{B}}$ in the
usual form
\begin{equation}
|\Psi_{\mathrm{AB}}\rangle =
\alpha_1 | 0_{\mathrm{A}} 0_{\mathrm{B}} \rangle
+ \alpha_2 | 0_{\mathrm{A}} 1_{\mathrm{B}} \rangle
+ \alpha_3 | 1_{\mathrm{A}} 0_{\mathrm{B}} \rangle
+ \alpha_4 | 1_{\mathrm{A}} 1_{\mathrm{B}} \rangle ,
\end{equation}
where $\sum_j |\alpha_j|^2 =1$. However, for our
purposes here it proves more convenient to write
this state in terms of a Schmidt decomposition~\cite{Per95}
\begin{equation}
\label{genS}
|\Psi_{\mathrm{AB}} \rangle =
\kappa_1 |x_1, y_1 \rangle+
\kappa_2 |x_2, y_2 \rangle ,
\end{equation}
where $\{ |x_1 \rangle, | x_2 \rangle \}$
and $\{ |y_1 \rangle, | y_2 \rangle \}$
are orthonormal bases of $\mathcal{H}_{\mathrm{A}}$
and $\mathcal{H}_{\mathrm{B}}$, respectively,
and $\kappa_1$ and $\kappa_2$ are real
nonnegative coefficients satisfying
$\kappa_1^2 + \kappa_2^2 = 1$ and
$\kappa_1 \ge \kappa_2$. In particular,
we shall choose
\begin{eqnarray}
|x_k\rangle & = & a_k |0_{\mathrm{A}} \rangle
+ b_{k}|1_{\mathrm{A}} \rangle , \nonumber \\
|y_k\rangle & = & \alpha_k |0_{\mathrm{B}} \rangle
+ \beta_k |1_{\mathrm{B}} \rangle ,
\end{eqnarray}
for $k = 1,2$; i.e., the corresponding bases
are related by general local unitary
transformations (obviously, the coefficients
must fulfill constraints in order to ensure
the orhonormality of the transformed bases).

Let us assume that the phases of this bipartite
system are described by a POVM such as (\ref{Laque}).
For the general state described in (\ref{genS})
one can easily compute the joint probability
distribution $\mathcal{P} (\varphi_+, \varphi_-)$ as
\begin{eqnarray}
\mathcal{P} (\varphi_+ , \varphi_-) & = &
\frac{1}{(2 \pi )^2}
\sum_{k,l} \kappa_k \kappa_l [ 1 +
\gamma_{\mathrm{A}} \gamma_{\mathrm{B}}
\nonumber \\
& \times &  ( a_k \alpha_k b_l^\ast \beta_l^\ast
e^{i \varphi_+} +  b_k \beta_k
a_l^\ast \alpha_l^\ast e^{-i\varphi_+} +
a_k \beta_k b_l^\ast \alpha_l^\ast e^{i\varphi_-} +
\alpha_k b_k a_l^\ast \beta_l^\ast
e^{-i\varphi_-} ) ] .
\end{eqnarray}

Once obtained this joint probability distribution,
we can proceed further by calculating the
associated dispersions
\begin{eqnarray}
D^+ & = & 1 - \left | \int_{2 \pi} d \varphi_+ \
e^{\pm i \varphi_+} \mathcal{P} (\varphi_+,
\varphi_-) \right |^2  \nonumber \\
& = & 1- \left ( \frac{\gamma_{\mathrm{A}}
\gamma_{\mathrm{B}}}{4 \pi^2} \right )^2
\left |\sum_k \kappa_k a_k \alpha_k \right |^2
\left |\sum_l \kappa_l b_l \beta_l \right |^2 ,
\nonumber \\
D^- & = & 1 - \left | \int_{2 \pi} d \varphi_- \
e^{\pm i \varphi_-} \mathcal{P} (\varphi_+,
\varphi_-) \right |^2  \nonumber \\
& = & 1 - \left ( \frac{\gamma_{\mathrm{A}}
\gamma_{\mathrm{B}}}{4 \pi^2} \right )^2
\left |\sum_k \kappa_k \alpha_k b_k \right |^2
\left |\sum_l \kappa_l a_l \beta_l \right |^2 .
\end{eqnarray}
We stress that these are the only relevant
phase-related quantities involved in the
problem and, in addition, they are accessible
to the experiment~\cite{Jul01}.

We propose to define the degree of entanglement
by
\begin{equation}
\mathbb{D} = \frac{| D^+ - D^- |}{\Gamma} ,
\end{equation}
where the constant
\begin{equation}
\Gamma = \left ( \frac{\gamma_{\mathrm{A}}
\gamma_{\mathrm{B}}}{2 \pi^2} \right )^2
\end{equation}
has been chosen so as to normalize
$0 \le \mathbb{D} \le 1$. Obviously,
for any separable state we have $\kappa_1 =1,
\kappa_2 = 0$ and then $\mathbb{D} = 0$. On
the opposite limit, for the ``magic" Bell basis,
which is especially germane for displaying
correlations between Alice's and Bob's
qubits~\cite{Per95},
\begin{eqnarray}
| \Psi^\pm \rangle & = & \frac{1}{\sqrt{2}}
(|0_{\mathrm{A}} 1_{\mathrm{B}}  \rangle
\pm | 1_{\mathrm{A}} 0_{\mathrm{B}} \rangle ) ,
\nonumber \\
| \Phi^\pm \rangle & = & \frac{1}{\sqrt{2}}
(|0_{\mathrm{A}} 0_{\mathrm{B}} \rangle
\pm |1_{\mathrm{A}} 1_{\mathrm{B}} \rangle ) ,
\end{eqnarray}
one can immediately check that $\mathbb{D} = 1$.

As a simple, but nontrivial example, let us
consider the family of nonmaximally entangled
states parametrized by
\begin{equation}
|\Phi_\varepsilon^\pm \rangle = \frac{1}{\sqrt{2}}
\{ [ \varepsilon |0_{\mathrm{A}} \rangle
+ (1- \varepsilon) |1_{\mathrm{A}} \rangle]/\mathcal{N}
\otimes |0_{\mathrm{B}} \rangle
\pm |1_{\mathrm{A}} 1_{\mathrm{B}} \rangle \} ,
\end{equation}
where $\mathcal{N}$ is a constant ensuring
the proper normalization of the state. Note
that when $\varepsilon \rightarrow 1$ these
states become the (maximally entangled) Bell
states $| \Phi^\pm \rangle$, while for $\varepsilon
\rightarrow 0$ they tend to the (separable) states
$|1_{\mathrm{A}} \rangle (|0_{\mathrm{B}} \rangle
\pm |1_{\mathrm{B}})/\sqrt{2}$. For this family
one has
\begin{equation}
\mathbb{D} = \frac{\varepsilon^2}{\mathcal{N}^2} .
\end{equation}

Apart from mathematical subtleties, we think
that the appeal of this new measure is that
it relies on observable quantities that can
be recast in terms of measurements of two-particle
visibility~\cite{Jae93}. It is a general belief~\cite{Bou00}
that the information encoded by these bipartite
entangled states only lies in the relative
qubit properties and not in the local
ones. This is exactly the purpose of the
parameter $\mathbb{D}$: using  phase sum
and phase difference this property is fully
brought out.

\section{Conclusions}

In this paper we have investigated a description of the
phase for a qubit in terms of a proper polar decomposition
of its amplitude, much in the spirit of our previous
work on the subject. Perhaps, the most striking consequence
of this description is that such a phase can take only
two values: $\pm \pi/2$.

We have also considered some other generalized descriptions
in terms of POVMs. Although these formalisms give different
results, they share a lot of properties. In particular, we
have shown an effective discreteness even if, in principle,
a continuous range of variation is assumed.

We have discussed the subtleties that arise when considering
the relative phase for two qubits. We have presented a
procedure to cast individual phases to phase sum and
phase difference. The relative phase that emerges from this
casting procedure is a powerful tool for examining
entanglement. In particular, we have proposed a measure
of entanglement involving only relative-phase dispersions,
which are physically measurable quantities.

\end{document}